\documentclass[11pt,english]{revtex4}
\usepackage{lmodern}
\usepackage[T1]{fontenc}
\usepackage[latin9]{inputenc}
\usepackage{xcolor}
\usepackage{babel}
\usepackage{textcomp}
\usepackage{amsmath}
\usepackage{amssymb}
\usepackage{graphicx}
\usepackage[unicode=true]
 {hyperref}

\makeatletter

\@ifundefined{textcolor}{}
{%
 \definecolor{BLACK}{gray}{0}
 \definecolor{WHITE}{gray}{1}
 \definecolor{RED}{rgb}{1,0,0}
 \definecolor{GREEN}{rgb}{0,1,0}
 \definecolor{BLUE}{rgb}{0,0,1}
 \definecolor{CYAN}{cmyk}{1,0,0,0}
 \definecolor{MAGENTA}{cmyk}{0,1,0,0}
 \definecolor{YELLOW}{cmyk}{0,0,1,0}
}

\usepackage{enumerate}
\usepackage{fbox}
\usepackage{lettrine}

\makeatother

\begin{document}
\title{\noindent \emph{\normalsize{}Honorable Mention --- Essay written
for the Gravity Research Foundation\emph{\linebreak} 2023 Awards
for Essays on Gravitation}\vskip24pt Buchdahl-inspired spacetimes
and wormholes:\vskip2pt Unearthing Hans Buchdahl's other `hidden'
treasure trove\vskip4pt}
\author{Hoang Ky Nguyen$\,$}
\email[\ \ ]{hoang.nguyen@ubbcluj.ro}

\affiliation{Department of Physics, Babe\c{s}--Bolyai University, Cluj-Napoca
400084, Romania}
\date{March 27, 2023}
\begin{abstract}
\noindent Circa 1962 $\,$Hans A. Buchdahl pioneered a program --
and made significant progress -- seeking vacuo configurations in
pure $\mathcal{R}^{2}$ gravity (H.$\,$A.$\,$Buchdahl, \href{http://link.springer.com/article/10.1007/BF02733549}{Nuovo Cimento 23, 141 (1962)}
\citep{Buchdahl-1962}).\linebreak Unfortunately, he deemed the final
step in his calculations impassable and prematurely suspended his
pursuit. Since then, his achievements on this front have faded into
dormancy. Unbeknownst to Buchdahl himself, the $\mathcal{R}^{2}$
vacua that he sought were within \emph{his} striking\linebreak distance.
In our recent three-paper body of work published in Phys.$\,$Rev.$\,$D
\citep{Nguyen-2022-Buchdahl,Nguyen-2022-Lambda0,Nguyen-2022-extension},
we broke this six-decades-old impasse and accomplished his goal: $\,$A
novel class of metrics, describing non-Schwarzschild spacetimes in
quadratic gravity and carrying a hallmark of higher-derivative characteristic.
Intriguing properties of Buchdahl-inspired spacetimes and their associated
Morris-Thorne-Buchdahl wormholes -- summarized herein -- embody
a new branch of phenomenology that transcends the Einstein--Hilbert
paradigm.
\end{abstract}
\maketitle
\begin{flushright}
\emph{\vskip8ptMay this unsung legacy of Hans A. Buchdahl}\linebreak
\emph{long `forsaken' and `forgotten'}\linebreak \emph{finally live
on}\vskip30pt
\par\end{flushright}

\linespread{1.25}\selectfont{}

\noindent Our essay is a passageway bridging the past and the future.
The past is rooted in a research program originated  but left unfinished
 by Buchdahl going back threescore years ago. The future is the new
horizons his program brings forth. In between is our labor -- summarized
in a three-installment ``\emph{Beyond Schwarzschild--de Sitter spacetimes}''
series \citep{Nguyen-2022-Buchdahl,Nguyen-2022-Lambda0,Nguyen-2022-extension}
-- to advance the Buchdahl program and achieve his ultimate goal.\vskip4pt
\begin{center}
-----------------$\infty$-----------------
\par\end{center}

\textbf{\emph{Introduction.}}---Analytical solutions in the study
of gravitation are few and far between. Yet, one such solution with
intriguing theoretical appeals has escaped the attention of the gravitation
research community in the past sixty years.\vskip4pt

In this essay we shall unveil a new set of vacua for the pure $\mathcal{R}^{2}$
action in 3+1 dimensions. The great surprise is that the vacua were
already inherent in an insightful -- but now largely forgotten --
work \citep{Buchdahl-1962}, initiated by Buchdahl circa 1962 in search
of vacuo configurations for pure $\mathcal{R}^{2}$ gravity, a theory
considered back in his time to be a viable prototype of alternatives
to General Relativity (GR). Despite making an impressive headway,
in an unfortunate twist of events, Buchdahl deemed the final technical
hurdle in his program insurmountable and abandoned his pursuit for
an analytical solution. Lacking further tangible progress \citep{foot-1},
this work of Buchdahl on pure $\mathcal{R}^{2}$ gravity has also
been overshadowed by his other important `1970 paper that pioneered
the $f(R)$ action \citep{Buchdahl-1970}. Unbeknownst to Buchdahl,
the analytical solution -- that we shall report herein -- was within
\emph{his} reach. In a recent Phys.$\,$Rev.$\,$D publication \citep{Nguyen-2022-Buchdahl}
we managed to revitalize Buchdahl's program, bridge its final remaining
gap, and succeed in accomplishing his goal. The outcome is an exhaustive
class of vacuo metrics -- which we called Buchdahl-inspired metrics
-- in an exact compact form, describing spacetimes with \emph{non-constant}
scalar curvature and holding promise toward new physics.\vskip4pt

There is yet another crucial advance we have made in this direction.
In a follow-up paper \citep{Nguyen-2022-Lambda0} we identified a
particular member, within the class of Buchdahl-inspired metrics,
that admits an \emph{exact closed analytical} form and describes asymptotically
flat spacetimes. \vskip4pt

The importance of these vacua is multi-faceted. Firstly, they are
\emph{bona fide} enlargements of the Schwarzschild-de Sitter (SdS)
metric, which is a cornerstone of gravitation research, alongside
the Friedmann-Lema\^itre-Robertson-Walker (FLRW) and Kerr metrics,
due to their physical relevance and mathematical appeals. Although
exact solutions have been discovered for various alternative theories
of gravitation, most of them belonging to the $f(\mathcal{R})$ family,
they often face difficulties in passing the four classical tests of
GR \citep{ModGravReviews}. However, since our solutions constitute
a spectrum of metrics that supersedes the Schwarzschild metric, they
are highly likely to reproduce the phenomenology of GR that has been
established to date.\vskip4pt

Secondly, these vacua have the advantage, like the well-known trio
of flagship metrics -- the SdS, FLRW, and Kerr metrics -- of being
expressible in analytical forms. As such, they can serve as a benchmark
for the study and exploration of modified theories of gravity.\vskip4pt

Thirdly, due to the fourth-order structure of the $\mathcal{R}^{2}$
theory, our solutions possess a novel higher-derivative characteristic
that holds the potential to produce new phenomenology beyond the scope
of existing tests.\vskip4pt

Furthermore, this new phenomenology emerges from the parsimonious
pure $\mathcal{R}^{2}$ theory which does not require ``exotic'' fancy
ingredients, such as supplemental terms, torsion, non-metricity, or
non-locality. As a classical field theory, pure $\mathcal{R}^{2}$
gravity is known to be scale-invariant and free of ghosts -- desirable
features for exploring the quantum gravity realm \citep{AlvarezGaume-2015,Alvarez-2018,Stelle}.
In addition, pure $\mathcal{R}^{2}$ gravity propagates massless spin-2
and massless spin-0 excitations on a de Sitter background \citep{AlvarezGaume-2015}.
Recent years have also seen the introduction of extensions to the
theory that naturally incorporate the matter sector \citep{agravity}.
\begin{center}
-----------------$\infty$-----------------
\par\end{center}

\textbf{\emph{The Buchdahl program}}\textbf{.}---In a 1962 Nuovo
Cimento paper \citep{Buchdahl-1962}, Buchdahl initiated a program
to seek static spherically symmetric vacuo configurations for the
quadratic action, $\frac{1}{2\kappa}\int d^{4}x\sqrt{-g}\,\mathcal{R}^{2}$,
with $\mathcal{R}$ being the Ricci scalar. The $\mathcal{R}^{2}$
field equation in vacuo reads
\begin{equation}
\mathcal{R}\Bigl(\mathcal{R}_{\mu\nu}-\frac{1}{4}g_{\mu\nu}\mathcal{R}\Bigr)+(g_{\mu\nu}\,\square-\nabla_{\mu}\nabla_{\nu})\mathcal{R}=0\label{eq:field-eqn}
\end{equation}
It automatically accepts (i) the null-Ricci-scalar spaces, $\mathcal{R}=0$,
and (ii) the Einstein spaces, $\mathcal{R}_{\mu\nu}=\Lambda\,g_{\mu\nu}$,
as solutions. The classic Schwarzschild-de Sitter (SdS) metric belongs
to the latter. As a fourth-order theory, however, the action is expected
to admit \emph{additional} solutions over and above the Einstein spaces.\vskip4pt

Buchdahl started with a judicious choice for the metric in a radial
coordinate $u$:
\begin{equation}
ds^{2}=-A(u)\,dt^{2}+B(u)\,du^{2}+\sqrt{\frac{B(u)}{A(u)}}\left(d\theta^{2}+\sin^{2}\theta\,d\varphi^{2}\right)
\end{equation}
In this \emph{harmonic gauge}, the trace of the vacuo field equation
\eqref{eq:field-eqn} is trivial, per
\begin{equation}
\square\,\mathcal{R}=0\ \ \ \Rightarrow\ \ \ \frac{d^{2}\mathcal{R}}{du^{2}}=0\ \ \ \Rightarrow\ \ \ \mathcal{R}=\Lambda+k\,u
\end{equation}
which readily yields a \emph{non-constant} scalar curvature \citep{foot-2}.
He then devised a sequence of mathematical maneuvers by introducing
a new radial coordinate $r$ and an auxiliary variable $q$ as an
undetermined function of $r$, with $A(u)$ and $B(u)$ expressed
in relation to $q(r)$. The entire endeavor was reduced to a non-linear,
second-order ordinary differential equation (ODE) of the function
$q(r)$, which we called \emph{the Buchdahl equation} in his honor
\citep{foot-3}:
\begin{equation}
\frac{d^{2}q}{dr^{2}}+\left(\frac{2\,\Lambda\,r}{1-\Lambda\,r^{2}}-\frac{3\,k^{2}}{4\,r}\frac{1}{q^{2}}\right)\frac{dq}{dr}=0\label{eq:Buchdahl-ODE}
\end{equation}
\emph{If an analytical solution for $q(r)$ to this ODE can be found,
then the vacuo he sought would ensue}. Unfortunately, Buchdahl deemed
Eq. \eqref{eq:Buchdahl-ODE} intractable and ceased his exploration.\vskip8pt

\begin{figure}[t]
\noindent \begin{raggedleft}
{\small{}\includegraphics[scale=0.62]{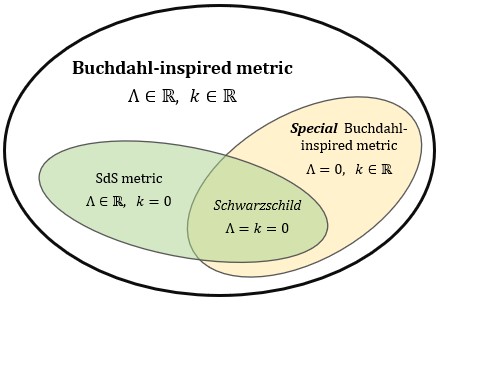}\includegraphics[scale=0.62]{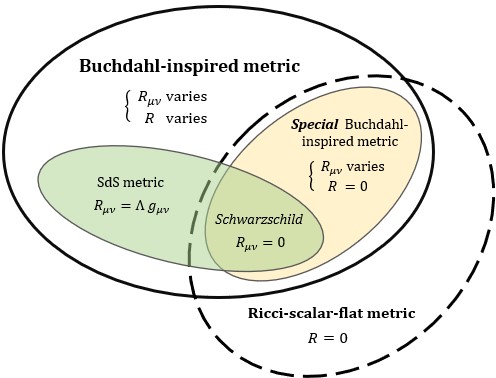}}{\small\par}
\par\end{raggedleft}
{\small{}\caption{\label{fig:Buchdahl-inspired-metric-family}$\ $Left plot: the Buchdahl-inspired
metric family and its subsets. Right plot: their relation with the
Ricci-scalar-flat family. The \emph{special} Buchdahl-inspired metric
occupying the intersection is asymptotically flat; the Buchdahl-inspired
metric with $\Lambda\protect\neq0$ is asymptotically de Sitter.}
}{\small\par}
\end{figure}

\textbf{\emph{The generic Buchdahl-inspired vacuo solution.}}---In
Ref. \citep{Nguyen-2022-Buchdahl} we successfully overcame this longstanding
impasse of over six decades and achieved Buchdahl's objective. We
tackled the Buchdahl equation, Eq.$\,$\eqref{eq:Buchdahl-ODE}, resulting
in a precise and concise expression for a comprehensive class of metrics
that are defined by the two higher-derivative characteristics, $\Lambda$
and $k$ (the latter being named \emph{the Buchdahl parameter}), cast
in the following form
\begin{equation}
\addtolength{\fboxsep}{10pt}\boxed{ds^{2}=e^{k\int\frac{dr}{r\,q(r)}}\ \left\{  \,p(r)\,\Bigl[-\frac{q(r)}{r}\,dt^{2}+\frac{r}{q(r)}\,dr^{2}\Bigr]+r^{2}\,d\Omega^{2}\right\}  }\label{eq:B-metric}
\end{equation}
The two auxiliary variables $p(r)$ and $q(r)$ obey a coupled set
of ``evolution'' rules
\begin{equation}
\frac{dp}{dr}=\frac{3\,k^{2}}{4\,r}\frac{p}{q^{2}}\ \ \ \ \ \text{and}\text{\ \ \ \ \ }\frac{dq}{dr}=\left(1-\Lambda\,r^{2}\right)p\label{eq:evolution}
\end{equation}
whereas the Ricci scalar is given by
\begin{equation}
\mathcal{R}(r)=4\Lambda\,e^{-k\int\frac{dr}{r\,q(r)}}\label{eq:B-Ricci}
\end{equation}
In order to firmly establish its validity, we directly inspected and
verified that \emph{the Buchdahl-inspired metric}, given in Eqs. \eqref{eq:B-metric}
and \eqref{eq:evolution}, satisfies the vacuo field equation \eqref{eq:field-eqn}
for all $\Lambda\in\mathbb{R}$ and all $k\in\mathbb{R}$. As per
Eq. \eqref{eq:B-Ricci}, the metric has a \emph{non-constant} scalar
curvature as long as $\Lambda\neq0$ and $k\neq0$. When $k=0$, the
evolution rules \eqref{eq:evolution} result in the solution: $p=1$,
$q=r-r_{\text{s}}-\frac{\Lambda}{3}r^{3}$, reproducing the SdS metric
with a constant Ricci scalar of $4\Lambda$. Therefore, the Buchdahl-inspired
metric family is a \emph{bona fide} enlargement of the SdS metric,
with their relationship illustrated in Fig. \ref{fig:Buchdahl-inspired-metric-family}.
Due to the nonlinear interplay between $p(r)$ and $q(r)$ in \eqref{eq:evolution},
the metric displays a diverse range of behaviors, which will be presented
in our report \citep{Nguyen-2023-Properties}.{\small{}}
{\small\par}
\begin{center}
-----------------$\infty$-----------------
\par\end{center}

\textbf{\emph{The special Buchdahl-inspired metric.}}\emph{---}Surprisingly,
despite their non-linearity, the evolution rules in \eqref{eq:evolution}
are \emph{fully soluble} for $\Lambda=0$. The detailed derivation
of this metric -- which we called \emph{the special Buchdahl-inspired
metric} -- is given in a companion report \citep{Nguyen-2022-Lambda0}.
Therein, by solving \eqref{eq:evolution} with $\Lambda=0$, we obtained
the following solution
\begin{align}
r & =\left|q-q_{+}\right|^{\frac{q_{+}}{q_{+}-q_{-}}}\left|q-q_{-}\right|^{-\frac{q_{-}}{q_{+}-q_{-}}}\label{eq:r-vs-q}\\
p & =\frac{(q-q_{+})(q-q_{-})}{r\,q}\\
q_{\pm} & :=\frac{1}{2}\Bigl(-r_{\text{s}}\pm\sqrt{r_{\text{s}}^{2}+3k^{2}}\Bigr)
\end{align}
with $r_{\text{s}}$ being an integration constant and $q_{\pm}$
the two real roots of the algebraic equation $q^{2}+r_{\text{s}}q-\frac{3k^{2}}{4}=0$.
Interestingly, although the Buchdahl ODE \eqref{eq:Buchdahl-ODE}
describes $q$ as a function of $r$, the solution \eqref{eq:r-vs-q}
has $r$ expressed in terms of $q$.\vskip4pt

We then made the following transformation from $r$ to a \emph{new}
radial coordinate $\rho$:\vskip1pt{\small{}
\begin{equation}
r(\rho):=\frac{\zeta\,r_{\text{s}}\left|1-\frac{r_{\text{s}}}{\rho}\right|^{\frac{1}{2}(\zeta-1)}}{1-\text{sgn}\bigl(1-\frac{r_{\text{s}}}{\rho}\bigr)\left|1-\frac{r_{\text{s}}}{\rho}\right|^{\zeta}}\ \ \ \ \ \text{with}\ \ \ \ \ \zeta:=\sqrt{1+\frac{3k^{2}}{r_{\text{s}}^{2}}}\label{eq:r-vs-rho}
\end{equation}
}\vskip1ptIn this new coordinate, the special Buchdahl-inspired metric
takes on a strikingly well-structured form. It is specified by a ``Schwarzschild''
radius $r_{\text{s}}$ and \emph{the scaled (dimensionless) Buchdahl
parameter} $\tilde{k}:=k/r_{\text{s}}$ per\vskip-8pt

\begin{equation}
\addtolength{\fboxsep}{10pt}\boxed{ds^{2}=\left|1-\frac{r_{\text{s}}}{\rho}\right|^{\tilde{k}}\ \left\{  -\Bigl(1-\frac{r_{\text{s}}}{\rho}\Bigr)\,dt^{2}+\frac{d\rho^{2}}{1-\frac{r_{\text{s}}}{\rho}}\frac{r^{4}(\rho)}{\rho^{4}}+r^{2}(\rho)\,d\Omega^{2}\right\}  }\label{eq:special-B-metric}
\end{equation}

\vskip6ptFor $\tilde{k}\neq0$, the special Buchdahl-inspired metric
is not Ricci-flat and is therefore non-Schwarzschild. However, it
is asymptotically flat and Ricci-scalar-flat. When $\tilde{k}=0$,
it precisely recovers the Schwarzschild metric as $r(\rho)\equiv\rho$
for all $\rho\in\mathbb{R}$. The relationships between the special
Buchdahl-inspired metric and others are depicted in Fig. \ref{fig:Buchdahl-inspired-metric-family}.\vskip8pt

\emph{Kretschmann invariant.}---The Kretschmann scalar, $K:=\mathcal{R}^{\mu\nu\rho\sigma}\mathcal{R}_{\mu\nu\rho\sigma}$,
of the special Buchdahl-inspired metric \eqref{eq:special-B-metric}
is given by (with $x$ shorthand for $1-r_{\text{s}}/\rho$)\vskip-6pt

{\footnotesize{}
\begin{align}
K & =2\zeta^{-8}r_{\text{s}}^{-4}\,\Bigl(1-\text{sgn}(x)\left|x\right|^{\zeta}\Bigr)^{6}\left|x\right|^{2-4\zeta-2\tilde{k}}\times\nonumber \\
 & \ \ \ \ \Bigl\{4\tilde{k}^{2}(\tilde{k}+1)\,\text{sgn}(x)\left|x\right|^{\zeta}+\bigl(-4\tilde{k}^{3}+5\tilde{k}^{2}+3\zeta\bigr)\bigl(\left|x\right|^{2\zeta}-1\bigr)+\bigl(9\tilde{k}^{4}-2\tilde{k}^{3}+10\tilde{k}^{2}+3\bigr)\bigl(\left|x\right|^{2\zeta}+1\bigr)\Bigr\}\label{eq:Kretschmann}
\end{align}
}The Kretschmann scalar exhibits a (curvature) singularity on the
boundary between the interior and exterior regions, namely $\rho=r_{\text{s}}$,
except for the cases where $\tilde{k}=0$ and $\tilde{k}=-1$. Furthermore,
the singularity at the origin, $\rho=0$, remains present for all
$k\in\mathbb{R}$.\vskip8pt
\begin{center}
-----------------$\infty$-----------------
\par\end{center}

\begin{figure}[!t]
\noindent \begin{centering}
\includegraphics[scale=0.72]{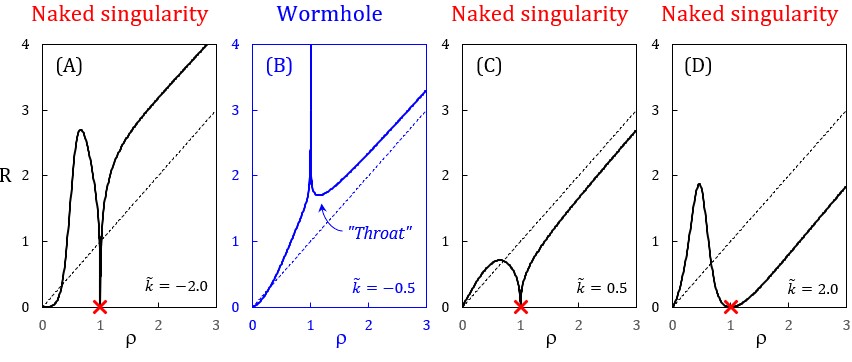}
\par\end{centering}
\caption{\label{fig:R-vs-rho}$\ $Behavior of $R(\rho)$ for metric \eqref{eq:special-B-metric},
with $r_{\text{s}}=1$. Panel (B), representative of $\tilde{k}\in(-1,0)$,
displays a minimum for $R(\rho)$ in the \textquotedblleft exterior\textquotedblright{}
and corresponds to a wormhole \textquotedblleft throat\textquotedblright .
The three remaining panels, representative of $\tilde{k}\in(-\infty,-1)\cup(0,+\infty)$,
each shows a red cross indicating a naked singularity at $\rho=r_{\text{s}}$.}
\end{figure}

\textbf{\emph{Construction of Morris-Thorne-Buchdahl wormholes.}}\emph{---}The
areal radius, given by $R(\rho)=\left|1-\frac{r_{\text{s}}}{\rho}\right|^{\frac{\tilde{k}}{2}}r(\rho)$,
produces a minimum for $R$ in the exterior $\rho>r_{\text{s}}$,
if and only if
\begin{equation}
-1<\tilde{k}<0
\end{equation}
Figure \ref{fig:R-vs-rho} shows the function $R(\rho)$ plotted for
different values of $\tilde{k}$. Panels (A), (C), and (D), representative
of $\tilde{k}\in(-\infty,-1)\cup(0,+\infty)$, show a monotonic function
$R(\rho)$ in the exterior and a naked singularity at $\rho=r_{\text{s}}$.
On the other hand, Panel (B) represents $\tilde{k}\in(-1,0)$ and
displays a minimum at $\rho_{0}$ in the region $\rho>r_{\text{s}}$,
indicating the presence of a wormhole ``throat''. The exterior region
$\rho\geq\rho_{0}$, together with its mirror image in the $\zeta-$KS
diagram, are glued together at the ``throat'', $\rho=\rho_{0}$, to
form \emph{a (two-way) traversable wormhole} which connects two asymptotically
flat spacetime sheets. \vskip8pt

Additional intriguing properties of Morris-Thorne-Buchdahl wormholes
\citep{MorrisThorne}, including the violation of the Weak Energy
Condition, in the broader context of scalar-tensor gravity will be
presented in our paper \citep{Nguyen-2023-MTB}.\vskip8pt

\begin{center}
-----------------$\infty$-----------------
\par\end{center}

\textbf{\emph{The $\zeta-$Kruskal-Szekeres (KS) diagram}}.---The
standard KS diagram for the Schwarzschild metric \citep{KS} can be
generalized for the special Buchdahl-inspired metric. We find that
\emph{the $\zeta-$KS coordinates}, $T$ and $X$, satisfy
\begin{equation}
T^{2}-X^{2}=-\text{sgn(\ensuremath{\rho-r_{\text{s}})}}e^{\frac{\rho^{*}(\rho)}{r_{\text{s}}}}\ \ \ \ \ \text{and}\ \ \ \ \ \frac{T}{X}=\left(\tanh\frac{t}{2r_{\text{s}}}\right)^{\text{sgn(\ensuremath{\rho-r_{\text{s}})}}}\label{eq:T-X}
\end{equation}
in which \emph{the $\zeta-$tortoise coordinate} $\rho^{*}(\rho)$
is defined via $d\rho^{*}:=\frac{r^{2}(\rho)}{\rho^{2}(1-\frac{r_{\text{s}}}{\rho})}\,d\rho$.
Restricted to the radial direction, viz. $d\theta=d\phi=0$, in the
$\zeta-$KS coordinates $(T,\,X)$, the special Buchdahl-inspired
metric \eqref{eq:special-B-metric} accepts the form
\begin{equation}
ds^{2}=-4r_{\text{s}}^{2}e^{-\frac{\rho^{*}(\rho)}{r_{\text{s}}}}\left|1-\frac{r_{\text{s}}}{\rho}\right|^{1+\tilde{k}}\left(dT^{2}-dX^{2}\right)\label{eq:zeta-KS}
\end{equation}
which appropriately recovers the usual KS result for $\tilde{k}=0$
\citep{KS}.\vskip4pt
\begin{figure}[!t]
\noindent \begin{centering}
\includegraphics[scale=0.54]{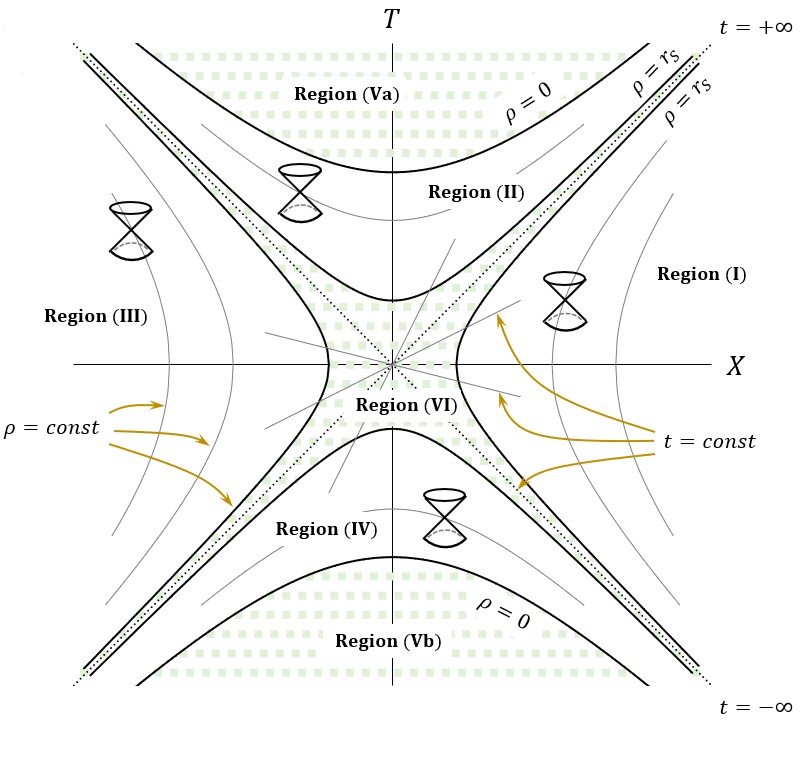}$\ \ \ \ \ \ $
\par\end{centering}
\caption{\label{fig:KS-diagram}$\ $The $\zeta-$Kruskal-Szekeres diagram
for the \emph{special} Buchdahl-inspired metric given by expressions
\eqref{eq:special-B-metric} and \eqref{eq:zeta-KS}. The \textquotedblleft gulf\textquotedblright{}
as Region (VI) is a new feature. See text for explanation.}
\end{figure}

Figure \ref{fig:KS-diagram} illustrates the $\zeta-$KS plane for
metric \eqref{eq:zeta-KS}, which is conformally Minkowski, analogous
to the standard KS diagram. The null geodesics, $dX=\pm dT$, align
on the $45^{\circ}$ lines in the $\zeta-$KS plane. Evidently, the
$\zeta$\textminus KS diagram preserves most of the causal structure
that has been established for the regular KS diagram. However, there
are some fundamental changes -- both qualitative and quantitative
-- to be described below.\vskip4pt

The boundary $\rho=r_{\text{s}}$ amounts to \emph{two} \emph{distinct}
hyperbolae, one for the interior and the other for the exterior, per
\begin{equation}
T^{2}-X^{2}=\begin{cases}
-e^{-\frac{\pi}{\sin\frac{\pi}{\zeta}}} & \text{for exterior}\\
+e^{-\frac{\pi}{\sin\frac{\pi}{\zeta}}} & \text{for interior}
\end{cases}\label{eq:hyperbolae}
\end{equation}
Note that each hyperbola comprises of two separate branches on its
own. For $\tilde{k}=0$, the hyperbolae \eqref{eq:hyperbolae} degenerate
to two straight lines $T=\pm X$ as expected for Schwarzschild black
holes.\vskip4pt

Region (I) represents the exterior of a black hole and is mapped onto
the $\zeta$--KS plane, extending up to the right branch of the hyperbola
$T^{2}-X^{2}=-e^{-\frac{\pi}{\sin\frac{\pi}{\zeta}}}$. Region (II)
denotes the interior of the black hole and is also mapped onto the
$\zeta$--KS plane, but this region extends up to the upper branch
of the hyperbola $T^{2}-X^{2}=+e^{-\frac{\pi}{\sin\frac{\pi}{\zeta}}}$.
Within Region (II), all null and timelike paths ultimately lead to
the origin, marked by the hyperbola with $\rho=0$, with no escape
possible. Outgoing light rays from Region (I) can escape to infinity,
while incoming light rays must enter Region (II) by ``bypassing''
Region (VI). Infalling objects hit the $\rho=r_{\ensuremath{\text{s}}}$
boundary on the side of Region (I) then reappears on the boundary
on the side of Region (II). Regions (III) and (IV) are the time-reversed
images of Regions (I) and (II), respectively. Regions (Va) and (Vb)
are unphysical, shaded in dots and representing $\rho<0$.\vskip4pt

Region (VI) is a new feature, also marked by dots, which sandwiches
between the four hyperbola branches in Eq. \eqref{eq:hyperbolae}.
Although it appears as a ``gulf'' in the $(T,\,X)$ coordinate system,
it does not correspond to any region in the original $(t,\,\rho)$
coordinate system. As $\tilde{k}$ approaches zero, the ``gulf'' shrinks
towards the lines $T=\pm X$, and it disappears when $\tilde{k}=0$.
By being \emph{the maximal analytical extension} of the special Buchdahl-inspired
metric, the $\zeta$--KS diagram is able to expose the peculiar Region
(VI) for this metric. The singularity of the Kretschmann invariant
on the interior-exterior boundary, the emergence of a traversable
Morris-Thorne-Buchdahl wormhole, and the ``gulf'' in the $\zeta$--KS
diagram collectively indicate that the topology of $\mathcal{R}^{2}$-gravity
spacetimes surrounding a mass source undergoes fundamental alterations
due to the Buchdahl parameter $\tilde{k}$, which is a higher-derivative
characteristic.
\begin{center}
-----------------$\infty$-----------------
\par\end{center}

\textbf{\emph{Conclusions.}}\emph{---}We revealed herein two novel
sets of vacua in pure $\mathcal{R}^{2}$ gravity, which is a compact
extension of GR. One set describes asymptotically de Sitter spacetimes,
while the other set describes asymptotically flat spacetimes. These
spacetimes were recently derived from Buchdahl's 1962 research program
\citep{Buchdahl-1962}, which had been largely overlooked until now.
They supersede the Schwarzschild-de Sitter spacetimes and point toward
potential far-reaching consequences that go beyond the Einstein-Hilbert
framework.\vskip4pt

It is remarkable that pure $\mathcal{R}^{2}$ gravity is a parsimonious
theory consisting of just one single term for the gravitation sector,
without the need for exogenous terms, torsion, non-metricity, metric-affine
hybrid, or non-locality, which are often included in modified theories
of gravity \citep{ModGravReviews}. This theory operates within the
vanilla local metric-based formalism. Yet, despite its simplicity,
it produces novel behaviors, as documented in this essay, which are
not encountered in the Einstein-Hilbert theory. Additionally, pure
$\mathcal{R}^{2}$ gravity allows for Buchdahl-inspired vacua that
feature \emph{non-constant} scalar curvature, as indicated in Eq.$\,$\eqref{eq:B-Ricci},
offering a sharp rebuttal to the generalized Lichnerowicz theorem
proposed in \citep{Lichnerowicz}. The asymptotic scalar curvature
$4\Lambda$ and the Buchdahl parameter $k$ are two \emph{endogenous}
degrees of freedom, which arise in a fourth-order theory, such as
pure $\mathcal{R}^{2}$ gravity, as opposed to a second-order theory,
such as the Einstein-Hilbert action. The Buchdahl parameter $k$ is
responsible for enriching $\mathcal{R}^{2}$ gravity with phenomenology
that transcends the Einstein-Hilbert paradigm.\vskip4pt

Supermassive objects should serve as test bed for theories of gravitation.
Our special Buchdahl-inspired solution, given in a closed analytical
form \eqref{eq:special-B-metric}, would naturally lend itself toward
stationary axisymmetric candidate solutions for rotating mass sources.
While progressing in this direction \citep{Nguyen-2023-axisym}, an
estimation for $\tilde{k}$ has been obtained through the M87 shadow,
resulting in the range of $-0.155\le\tilde{k}\le0.004$ which potentially
supports a wormhole. Our study of the Morris-Thorne-Buchdahl metric
\citep{Nguyen-2023-MTB} indicates that interpreting the supermassive
M87 as a wormhole is plausible within the context of pure $\mathcal{R}^{2}$
gravity. Typically, wormholes are supported by ``exotic'' matter that
violates the Weak Energy Condition \citep{WEC}. However, in pure
$\mathcal{R}^{2}$ gravity, the scalar degree of freedom that is associated
with the higher-order nature of the theory plays the role of exotic
matter without the need for truly exotic matter.\vspace{.2cm}
\begin{center}
-----------------$\infty$-----------------
\par\end{center}

\begin{center}
\textbf{\vspace{.25cm}}
\par\end{center}

\begin{center}
\textbf{\large{}Epilogue: $\,$A tribute to Buchdahl}\textbf{\vspace{1cm}}
\par\end{center}

\noindent \lettrine[findent=3pt, nindent=7pt, ante=\rlap{\color{white}\rule[-0.55\LettrineHeight]{\LettrineWidth}{0.65\LettrineHeight}}]{\textbf{H}}      {ans} {\scriptsize{}ADOLF
BUCHDAHL} departed the earthly world in 2010, having led an extraordinary
life and an illustrious career \citep{Goenner-2010}. Shortly after
his passing, I came across his `forsaken' treasure trove, left `hidden'
in plain sight for fifty odd years, and began my labor cutting and
polishing the gemstones... Fast forward ten more years. It is time
the physics world marveled at the jewels Buchdahl has passed on to
us. And our children... May his long `forgotten' intellectual offspring
finally shine through and open passages to new horizons.

\vspace{2.5cm}
\begin{center}
\textbf{\emph{Acknowledgments}}\emph{\vskip8ptI thank Dieter L\"ust,
Tiberiu Harko, Richard Shurtleff, Mustapha Azreg-A\"inou, Timothy
Clifton, Sergei Odintsov and Nicholas Buchdahl for their comments
or encouragements}
\par\end{center}

\linespread{1.3}\selectfont{}


\newpage
\begin{thebibliography}{1}
\bibitem{Buchdahl-1962}H.$\,$A. Buchdahl, \emph{On the Gravitational
Field Equations Arising from the Square of the Gaussian Curvature},
Nuovo Cimento \textbf{23}, 141 (1962), \textcolor{purple}{\href{http://link.springer.com/article/10.1007/BF02733549}{link.springer.com/article/10.1007/BF02733549}}

\bibitem{Nguyen-2022-Buchdahl}H.$\,$K. Nguyen, \emph{Beyond Schwarzschild-de
Sitter spacetimes: I. A new exhaustive class of metrics inspired by
Buchdahl for pure $R^{2}$ gravity in a compact form},\emph{ }Phys.$\,$Rev.$\,$D
\textbf{106}, 104004 (2022)\textcolor{purple}{, \href{https://arxiv.org/abs/2211.01769}{arXiv:2211.01769 [gr-qc]}}

\bibitem{Nguyen-2022-Lambda0}H.$\,$K. Nguyen, \emph{Beyond Schwarzschild-de
Sitter spacetimes: II. An exact non-Schwarzschild metric in pure $R^{2}$
gravity and new anomalous properties of $R^{2}$ spacetime}s, Phys.$\,$Rev.$\,$D
\textbf{10}7, 104008 (2023)\textcolor{purple}{,} \textcolor{purple}{\href{https://arxiv.org/abs/2211.03542}{arXiv:2211.03542 [gr-qc]}}

\bibitem{Nguyen-2022-extension} H.$\,$K. Nguyen, \emph{Beyond Schwarzschild-de
Sitter spacetimes: III. A perturbative vacuum with non-constant scalar
curvature in $R+R^{2}$ gravity}, Phys.$\,$Rev.$\,$D \textbf{10}7,
104009 (2023)\textcolor{purple}{,} \textcolor{purple}{\href{https://arxiv.org/abs/2211.07380}{arXiv:2211.07380 [gr-qc]}}

\bibitem{foot-1}Buchdahl's 1962 Nuovo Cimento paper \citep{Buchdahl-1962}
has gathered merely $40+$ citations to date.

\bibitem{Buchdahl-1970}H.$\,$A. Buchdahl, \emph{Nonlinear Lagrangians
and cosmological theory}, Mon.$\,$Not.$\,$Roy.$\,$Astron.$\,$Soc.
\textbf{150}, 1 (1970)

\bibitem{ModGravReviews}Excellent reviews are, e.g., T. Clifton,
P.$\,$G. Ferreira, A. Padilla, and C. Skordis, \emph{Modified gravity
and cosmology}, Phys.$\,$Rept. \textbf{513}, 1 (2012), \textcolor{purple}{\href{https://arxiv.org/abs/1106.2476}{arXiv:1106.2476 [astro-ph.CO]};}
A. De Felice and S. Tsujikawa, \emph{$f(R)$ theories}, Living Rev.$\,$Rel.
\textbf{13}, 3 (2010), \textcolor{purple}{\href{https://arxiv.org/abs/1002.4928}{arXiv:1002.4928 [gr-qc]};
}T.$\,$P. Sotiriou and V. Faraoni, \emph{$f(R)$ Theories Of Gravity},
Rev.$\,$Mod.$\,$Phys. \textbf{82}, 451 (2010), \textcolor{purple}{\href{https://arxiv.org/abs/0805.1726}{arXiv:0805.1726 [gr-qc]};}
S. Capozziello and M. De Laurentis, \emph{Extended Theories of Gravity},
Phys.$\,$Rept. \textbf{509}, 167 (2011), \textcolor{purple}{\href{https://arxiv.org/abs/1108.6266}{arXiv:1108.6266 [gr-qc]};
}S. Nojiri and S.$\,$D. Odintsov, \emph{Unified cosmic history in
modified gravity: from $F(R)$ theory to Lorentz non-invariant models},
Phys.$\,$Rept. \textbf{505}, 59 (2011), \textcolor{purple}{\href{https://arxiv.org/abs/1011.0544}{arXiv:1011.0544 [gr-qc]}}

\bibitem{AlvarezGaume-2015}L. Alvarez-Gaume, A. Kehagias, C. Kounnas,
D. L\"ust, and A. Riotto, \emph{Aspects of Quadratic Gravity}, Fortsch.$\,$Phys.
\textbf{64}, 176 (2016), \textcolor{purple}{\href{https://arxiv.org/abs/1505.07657}{arXiv:1505.07657 [hep-th]}}

\bibitem{Alvarez-2018}E. Alvarez, J. Anero, S. Gonzalez-Martin, and
R. Santos-Garcia, \emph{Physical content of quadratic gravity,} Eur.$\,$Phys.$\,$J.$\,$C
\textbf{78}, 794 (2018), \textcolor{purple}{\href{https://arxiv.org/abs/1802.05922}{arXiv:1802.05922 [hep-th]}}

\bibitem{Stelle}K.$\,$S. Stelle, \emph{Renormalization of higher-derivative
quantum gravity,} Phys.$\,$Rev.$\,$D \textbf{16}, 953 (1977); K.$\,$S.
Stelle, \emph{Classical Gravity with Higher Derivatives}, Gen.$\,$Rel.$\,$Grav.
\textbf{9}, 353 (1978)

\bibitem{agravity}A. Salvio and A. Strumia, \emph{Agravity}, J.$\,$High
Eneregy Phys. \textbf{06}, 080 (2014), \textcolor{purple}{\href{https://arxiv.org/abs/1403.4226}{arXiv:1403.4226 [hep-ph]}};
M.$\,$B. Einhorn and D.$\,$R. Timothy Jones, \emph{Grand Unified
Theories in renormalisable, classically scale invariant gravity},
J.$\,$High Energy Phys. \textbf{10}, 012 (2019)\textcolor{purple}{,
\href{https://arxiv.org/abs/1908.01400}{arXiv:1908.01400 [hep-th]}}

\bibitem{foot-2}Several existing works on pure $\mathcal{R}^{2}$
gravity missed out the Buchdahl-inspired solution as they solely sought
solutions that have constant scalar curvature. The oversight of these
works was due to the generalized Lichnerowicz theorem \citep{Lichnerowicz}
which, however, contains detrimental gaps; see our exposition of the
gaps in \citep{Nguyen-2022-extension}.

\bibitem{foot-3}Note that in his original derivation \citep{Buchdahl-1962},
Buchdahl missed out the $\Lambda$'s in Eq. \eqref{eq:Buchdahl-ODE}.
We were able to uncover them by reworking his derivation directly
from the $\mathcal{R}^{2}$ field equations; see \citep{Nguyen-2022-Buchdahl}.

\bibitem{Nguyen-2023-Properties}H.$\,$K. Nguyen, \emph{Properties
of Buchdahl-inspired metrics in $\mathcal{R}^{2}$ gravity} (to be
posted)

\bibitem{MorrisThorne}M.$\,$S. Morris and K.$\,$S. Thorne, \emph{Wormholes
in spacetime and their for interstellar travel: A tool for teaching
general relativity}, Am.$\,$J.$\,$Phys. \textbf{56}, 5 (1988); M.$\,$S.
Morris, K.$\,$S. Thorne, and U. Yurtsever, \emph{Wormholes, Time
Machines, and the Weak Energy Condition}, Phys.$\,$Rev.$\,$Lett.
\textbf{61}, 1446 (1988)

\bibitem{Nguyen-2023-MTB}H.$\,$K. Nguyen and M. Azreg-A\"inou,
\emph{Traversable Morris-Thorne-Buchdahl wormholes in quadratic gravity},
\textcolor{purple}{\href{https://arxiv.org/abs/2305.04321}{arXiv:2305.04321 [gr-qc]}}

\bibitem{KS}M.$\,$D. Kruskal, \emph{Maximal Extension of Schwarzschild
Metric}, Phys.$\,$Rev. \textbf{119}, 1743 (1960); P. Szekeres, \emph{On
the Singularities of a Riemannian Manifold}, Publicationes Mathematicae
Debrecen \textbf{7}, 285 (1960); A.$\,$S. Eddington, \emph{A Comparison
of Whitehead's and Einstein's Formulae}, Nature (London) \textbf{113},
192 (1924); D. Finkelstein, \emph{Past-Future Asymmetry of the Gravitational
Field of a Point Particle}, Phys.$\,$Rev. \textbf{110}, 965 (1958)

\bibitem{Lichnerowicz}The generalized Lichnerowicz theorem for quadratic
gravity was ``proved'' by various authors in the past 10 years.
However, there are detrimental gaps in the proofs which we recently
exposed in our paper \citep{Nguyen-2022-extension}. The generalized
Lichnerowicz theorem was popularized in: W. Nelson, \textcolor{black}{\emph{Static
solutions for fourth order gravity}}\textcolor{black}{, Phys.}$\,$\textcolor{black}{Rev.}$\,$\textcolor{black}{D
}\textbf{\textcolor{black}{82}}\textcolor{black}{, 104026 (2010)}\textcolor{purple}{,
\href{https://arxiv.org/abs/1010.3986}{arxiv:1010.3986 [gr-qc]};
}H. L\"u, A. Perkins, C.$\,$N. Pope, and K.$\,$S. Stelle, \emph{Black
holes in higher-derivative gravity,} Phys.$\,$Rev.$\,$Lett. \textbf{114},
171601 (2015), \textcolor{purple}{\href{https://arxiv.org/abs/1502.01028}{arxiv:1502.01028 [hep-th]};
}H. L\"u, A. Perkins, C.$\,$N. Pope, and K.$\,$S. Stelle, \emph{Spherically
symmetric solutions in higher-derivative gravity}, Phys.$\,$Rev.$\,$D
\textbf{92}, 124019 (2015), \textcolor{purple}{\href{https://arxiv.org/abs/1508.00010}{arXiv:1508.00010 [hep-th]};
}A. Kehagias, C. Kounnas, D. L\"ust, and A. Riotto, \emph{Black hole
solutions in $R^{2}$ gravity}, J.$\,$High$\,$Energy$\,$Phys. \textbf{05},
143 (2015), \textcolor{purple}{\href{https://arxiv.org/abs/1502.04192}{arxiv:1502.04192 [hep-th]}}

\bibitem{Nguyen-2023-axisym}M. Azreg-A\"inou and H.$\,$K. Nguyen,
\emph{A stationary axisymmetric candidate solution for pure $\mathcal{R}^{2}$
gravity}, \textcolor{purple}{\href{https://arxiv.org/abs/2304.08456}{arXiv:2304.08456 [gr-qc]}}

\bibitem{WEC}E-A. Kontou and K. Sanders, \emph{Energy conditions
in general relativity and quantum field theory}, Class. Quantum Grav.
\textbf{37}, 193001 (2020), \textcolor{purple}{\href{https://arxiv.org/abs/2003.01815}{arXiv:2003.01815 [gr-qc]}}

\bibitem{Goenner-2010}H. Goenner, \emph{Hans A. Buchdahl (7.7.1919--7.1.2010)},
General Relativity and Gravitation \textbf{42}, 1049 (2010)\textcolor{teal}{,
}\textcolor{purple}{\href{http://link.springer.com/content/pdf/10.1007/s10714-010-0965-9.pdf}{link.springer.com/content/pdf/10.1007/s10714-010-0965-9.pdf}}
\end{thebibliography}
\end{document}